\documentclass[a4paper]{article}

\usepackage{INTERSPEECH2020}
\usepackage[colorinlistoftodos]{todonotes}
\usepackage{siunitx}
\usepackage[T1]{fontenc}

\title{HiFi-GAN: High-Fidelity Denoising and Dereverberation \\ Based on Speech Deep Features in Adversarial Networks}
\name{Jiaqi Su$^{1,2}$, Zeyu Jin$^2$, Adam Finkelstein$^1$}
\address{
  $^1$Princeton University ~~~
  $^2$Adobe Research}
\email{$^1$\{jiaqis,af\}@princeton.edu ~ $^2$zejin@adobe.com~~~~~~~}

\begin{document}

\maketitle
\begin{abstract}
Real-world audio recordings are often degraded by factors such as noise, reverberation, and equalization distortion. This paper introduces HiFi-GAN, a deep learning method to transform recorded speech to sound as though it had been recorded in a studio. We use an end-to-end feed-forward WaveNet architecture, trained with multi-scale adversarial discriminators in both the time domain and the time-frequency domain. It relies on the deep feature matching losses of the discriminators to improve the perceptual quality of enhanced speech. The proposed model generalizes well to new speakers, new speech content, and new environments. It significantly outperforms state-of-the-art baseline methods in both objective and subjective experiments.

\end{abstract}


\section{Introduction}
Real-world recordings captured in natural spaces with consumer-grade devices typically contain a wide variety of noise, reverberation, and equalization distortion. Yet many applications would benefit from clean, high-quality recordings.
Researchers have developed a variety of 
methods to reduce noise~\cite{ephraim1984speech,scalart1996speech}, reduce reverberation~\cite{naylor2010speech,kinoshita2013reverb}, correct equalization~\cite{germain2016equalization}, or enhance speech for downstream tasks such as speech recognition~\cite{donahue2018exploring}. However the combined problem of denoising, dereverberation and equalization matching is not sufficiently addressed. The goal of this paper is to enhance inexpensive real-world recorded speech to sound like professional studio-quality recordings for single-channel audio recordings. 

Traditional signal processing methods, such as Wiener filtering~\cite{scalart1996speech}, weighted linear prediction error~\cite{nakatani2010speech} and non-negative matrix factorization~\cite{duan2012speech,kagami2018joint}, operate in the time-frequency domain utilizing prior knowledge of the spectral structure of speech. They generalize well across environmental conditions, but offer only modest reduction in noise and reverberation.
Modern approaches use machine learning to model the mapping from noisy reverberant signal to clean signal and afford substantially improved performance over traditional methods.
Methods based on spectra transform the spectrogram of a distorted input signal to match that of a target clean signal by estimating either a direct non-linear mapping from input to target~\cite{han2015learning,xu2015regression}, or a mask over the input~\cite{williamson2017speech, mack2018single}. They require ISTFT or other methods to obtain waveform from predictions, which often produce audible artifacts due to missing or mismatching phase.

Recent advances in neural network architectures enable mapping directly over the waveform, despite the dual challenges of high resolution coupled with dense temporal structure at many scales.
WaveNet~\cite{van2016wavenet} and its feed-forward variants for speech enhancement~\cite{rethage2018WaveNet, su2019perceptually} leverage dilated convolution to enable a large receptive field while retaining a relatively small number of parameters. Wave-U-Net~\cite{stoller2018wave} adapts the U-Net structure to the time domain to combine features at different time scales for speech separation and enhancement tasks~\cite{macartney2018improved, giri2019attention}. Waveform-based methods tend to produce fewer phase-induced artifacts but have distortions of their own. Our experiments also find them sensitive to training data and difficult to generalize to unfamiliar noises and reverberation.


Generative adversarial networks (GANs) improve authenticity of synthesized audio by incorporating an adversarial loss from a discriminator~\cite{binkowski2019high, kumar2019melgan}. The discriminator learns to
identify whether a given audio example is real or fake, so as to encourage the generator to better approximate the distribution of real data. While researchers have explored GANs on spectral features~\cite{donahue2018exploring, fu2019metricgan}, SEGAN~\cite{pascual2017segan} and following works \cite{phan2020improving, pascual2019towards} show early success of GAN operated on waveform for speech enhancement. They reduce artifacts over other waveform-based networks, but each has its own unique artifact that becomes more noticeable for more distorted input. 
%

Since human hearing is sensitive to incoherence in the signal induced by artifact, it is necessary to model human perception and use it as optimization objectives. 
MetricGAN~\cite{fu2019metricgan} and QualityNet~\cite{fu2019learning} directly optimize over differentiable approximations of objective metrics such as PESQ and STOI; they help to reduce the artifacts but not significantly, as the metrics correlate poorly with human perception at short distances. 
%
As another way to model perceptual similarity, deep feature loss utilizes feature maps learnt for recognition tasks (which can be viewed as machine perception) to approximate human perception for other tasks, for example denoising~\cite{germain2019speech}. 
However, this approach relies on a fixed feature space learned via a task unrelated to denoising, and thus can under-perform in subsequent application scenarios with differing sound statistics.

Given the shortcomings of the above methods, we propose HiFi-GAN that combines an end-to-end feed-forward WaveNet architecture with the idea of deep feature matching in adversarial training, operated on both the time domain and the time-frequency domain. We use a set of discriminators on the waveform sampled at different rates, as well as a discriminator on the mel-spectrogram. These discriminators jointly evaluate several aspects of the generated audio, thus improving its perceived quality. The deep feature matching losses can dynamically adapt to the task based on the feature maps of the discriminators, stabilize GAN training and hence enhance discrimination capability. 
In summary, our main contributions are:
(1)~a generic high-fidelity speech enhancement method for noise, reverberation and distortion that generalizes to new speakers, new speech content and new environments;
(2)~an adversarial training procedure using multi-scale multi-domain discriminators together with their deep feature matching losses in application to speech enhancement;
and (3)~objective and subjective evaluations on both the generic speech enhancement task and the benchmark denoising task, demonstrating that our approach significantly outperforms the state-of-the-art baseline methods.

\begin{figure*}[ht]
  \centering
  \includegraphics[width=0.99\textwidth]{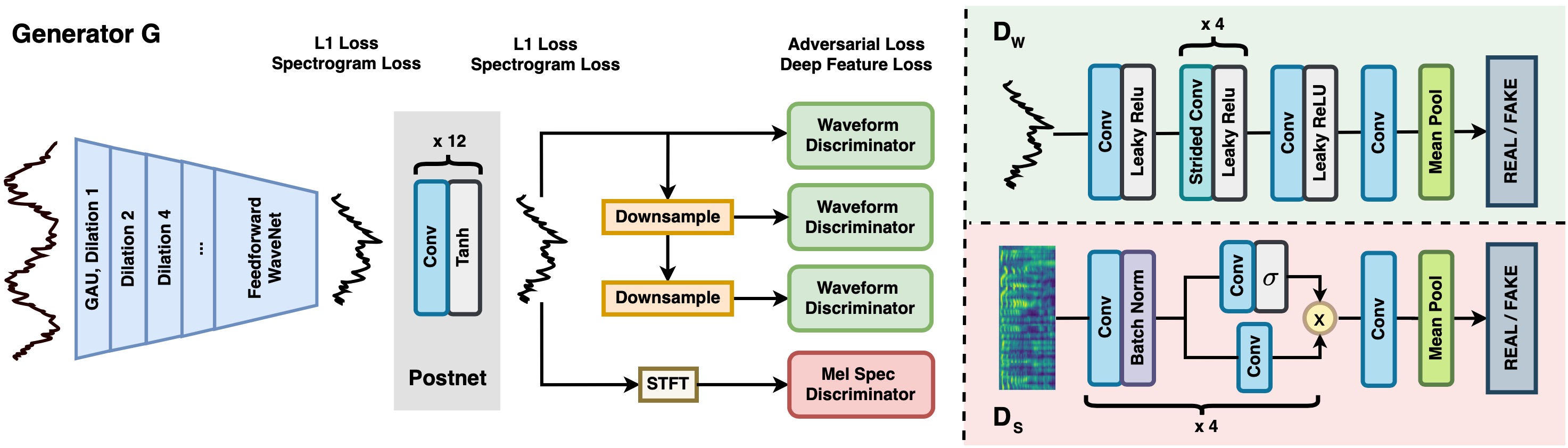}
   \vspace{-0.5\baselineskip}
  \caption{GAN Architecture. Generator G includes both a feed-forward WaveNet for speech enhancement, followed by a convolutional Postnet for cleanup. Discriminators evaluate the resulting waveform ($D_W$, at multiple resolutions) and mel-spectrogram ($D_S$).}
  \label{fig:network}
   \vspace{-0.5\baselineskip}
\end{figure*}
\section{Method}
Our method builds on top of our previous works of perceptually-motivated speech enhancement~\cite{su2019perceptually}. Though previous work aims at joint denoising and dereverberation on single recording environment, our goal is to generalize across environments. The overall architecture is shown in Figure~\ref{fig:network}. Our enhancement network (Generator G) is a feed-forward WaveNet~\cite{rethage2018WaveNet} which has shown success in denoising and dereverberation~\cite{qian2017speech, su2019perceptually}. Its non-casual dilated convolutions with exponentially increasing dilation rates enables a large receptive field suitable for typical additive noise and long tail reverberation. 
We also include log spectrogram loss and L1 sample loss as basic components of loss function, as they help to speed up training and serve as a simple perceptual metric. In practice, we use an equally weighted combination of two spectrogram losses for 16kHz audio: one with large FFT window size of 2048 and hop size of 512, and one with small FFT window size of 512 and hop size of 128. The larger one gives more frequency resolution, while the smaller one gives more temporal resolution. Our experiments in this paper operate on 16kHz to make it easy to compare with previous methods as they are developed at the same sample rate. However, true high-fidelity audio demands a higher sample rate, for which we discuss strategies in Section~\ref{sec:discussion}. Subsequent sections will discuss modifications to this architecture that bring out significant improvement in fidelity. 

\subsection{Postnet}
The use of postnets has been found effective in improving synthesis quality~\cite{shen2018natural}. We attach to the main network a simple stack of 12 1D-convolutional layers, each with 128 channels and kernel length 32. \texttt{Tanh} is used as activation function between convolutional layers. L1 and spectrogram losses are applied to both the output of the main network (before the postnet) and that after the postnet. This enables the main network to focus on generating coarse version of the clean speech while the postnet removes artifacts to improve fidelity.

\subsection{Adversarial training}
Additional adversarial training helps to expose subtle noise and artifacts that are not captured by simple loss functions. We employ both spectrogram and waveform discriminators to cover different domains and resolutions.
The generator is penalized with the adversarial losses as well as deep feature matching losses computed on the features maps of the discriminators. 

\subsubsection{Multi-scale multi-domain discriminators}
The design of discriminators is inspired by MelGAN~\cite{kumar2019melgan} which uses multi-scale discrimination on waveform in speech synthesis. Similarly, we use three waveform discriminators, respectively operating at 16kHz, 8kHz and 4kHz sampled versions of waveform, for discrimination at different frequency ranges. The waveform discriminators have the same network architecture but do not share the weights. Each network is composed of a 1D convolution, four strided convolution blocks, two additional 1D convolutions, and global average pooling to output a real-value score. Leaky Relu is used between the layers for non-linearity. Grouped convolution is used in place of traditional convolution to reduce the number of trainable parameters.
We found the waveform discriminators help to reduce noise and comb filtering artifact, enhancing speech clarity and speaker identity.

We also apply a discriminator on the mel-spectrogram, to tell generated speech from real speech. The L2 spectrogram loss often causes over-smoothing effects leading to increased aperiodicity in voiced sounds and ghosting effect due to ambiguity and mismatch between the F0 and partials. The spectrogram discriminator sharpens the corresponding spectrogram of predicted speech. It's also easier to identify long-span reverberation residual in the time-frequency domain than in the time domain. The network consists of four stacks of 2D convolution layer, batch normalization and Gated Linear Unit (GLU), similar to StarGAN-VC~\cite{kameoka2018stargan}

Having discriminators from two different domains stabilizes the training and balances the weighting of different factors of perceptual qualities, so that no single type of noise or artifact gets over-addressed. For all of our discriminators, we use the hinge version of the adversarial loss~\cite{lim2017geometric} for more efficient training. For a specific discriminator $D_{k}$, its adversarial loss on the generator $L_G^{\text{Adv}}$ and its discriminator loss $L_{D_{k}}$ are as below:
\begin{equation}
    L_G^{\text{Adv}} (x, x'; D_{k})=\max[1 - D_{k}(G(x)), 0]
  \label{eq1}
\end{equation}
\begin{equation}
\begin{aligned}
    L_{D_{k}} (x, x') &= \max[1 + D_{k}(G(x)), 0] \\
    &+\max[1 - D_{k}(x'), 0]
\end{aligned}
\label{eq2}
\end{equation}
where (x, x') is the pair of input audio $x$ and target audio $x'$.

In our experiments, we found that adversarial training works better together with the postnet than without. Our intuition is that it is easier for the discriminators to drive the postnet than the highly non-linear main network in order to reshape signal. While both improved with adversarial training, perceivable quality differences can be observed for outputs before the postnet and after the postnet.

\subsubsection{Deep feature matching loss}

Borrowed from Computer Vision~\cite{gatys2015neural}, the idea of deep feature loss has been applied to speech denoising~\cite{germain2019speech}, which uses a fixed feature space learnt from pre-training on tasks of environment detection and domestic audio tagging.
MelGAN~\cite{kumar2019melgan} shares a similar idea, computing matching loss on the deep feature maps of the intermediate layers in the discriminator during training. We apply it to speech enhancement to allow dynamically updating feature matching loss.
Since the discriminator is in dynamic competition with the generator, its deep feature matching loss stays on top to catch noticeable differences, enhancing the power of variable discrimination. Deep feature matching loss also prevents the generator from mode collapse, a common failure scenario of GANs~\cite{goodfellow2016nips}, by enforcing the generator to match to the reference content so that it can not cheat by producing monotonous examples.
Furthermore, it encourages the discriminator to learn a continuous feature space rather than memorizing the real samples, thus avoiding over-fitting. 

For a specific discriminator $D_{k}$, we formulate its deep feature matching loss on the generator as follows:
\vspace{-0.7\baselineskip}
\begin{equation}
    L_{G}^{\text{FM}} (x, x'; D_{k}) = \sum_{i=1}^{T_k} \frac{1}{N_{i}}||D_{k}^{(i)}(G(x)) - D_{k}^{(i)}(x')||_1 
\label{eq3}
\vspace{-0.5\baselineskip}
\end{equation}
where $T_k$ is the numebr of layers in $D_k$ excluding the output layer, and $N_{i}$ is the number of units in the $i$-th layer $D_k^{(i)}$.

To conclude, the constructed loss function on the enhancement network is a combination of L1 losses on waveforms (both before and after the postnet), L2 losses on log spectrograms (both before and after the postnet), adversarial losses $L_G^{\text{Adv}}$ and deep feature matching losses $L_G^{\text{FM}}$ from the four discriminators. This essentially can be viewed as a partially learnt perceptual loss function that evaluates various aspects of speech quality.

\subsection{Data simulation and augmentation}
The training of our model relies on parallel pairs of noisy recordings and clean recordings, which are expensive to obtain in reality. Instead, we create training data for various environments via data simulation by convolving studio quality recordings with measured room impulse responses and adding noise. 

To generalize to new speakers, new speech content, and new environments, we conduct heavy data augmentation on the fly~\cite{suacoustic}.
The amplitude and speed of speech is randomly scaled to add variation in speakers. Noise drawn from a sample collection is passed through a random multi-band filter and added with a random SNR between 10dB and 30dB. The amount of reverberation is augmented by adjusting the DRR and RT60 properties of the impulse responses following Bryan's proposed procedure~\cite{bryan2019data}. We also apply random multi-band filters to the impulse responses for equalization distortion. 

\section{Experiments}
Through experiments, our best performing generator network is a 20-layer feed-forward WaveNet with two stacks of dilated convolutions, each with filter size of 3 and dilation rates from 1 to 512. The channel size is 128 across the network. 
The spectrogram discriminator follows the same configuration as in StarGAN-VC~\cite{kameoka2018stargan}: kernel sizes of (3, 9), (3, 8), (3, 8), (3, 6); stride sizes of (1, 2), (1, 2), (1, 2), (1, 2); and channel sizes of 32 across the layers.
The input is computed as the 80-coefficient log mel-spectrogram with mels spanning from 20Hz to 8000Hz.
The waveform discriminators follow the same configuration as in MelGAN~\cite{kumar2019melgan}: 
kernel sizes of 15, 41, 41, 41, 41, 5, 3; stride sizes of 1, 4, 4, 4, 4, 1, 1; channel sizes of 16, 64, 256, 1024, 1024, 1024, 1; and group sizes of 1, 4, 16, 64, 256, 1, 1 for its grouped convolutions.

Training happens in three stages. First, we train the feed-forward WaveNet for 500K steps with learning rate 0.001, using L1 loss and spectrogram loss. This is to make sure the main network grasps the waveform structure of speech. Then we train together with the postnet and data augmentation for 500K steps, with learning rate 0.0001. Finally, we train the generator at learning rate 0.00001 with the four randomly initialized discriminators at learning rate 0.001 for 50K steps, using both adversarial losses and deep feature matching losses. We update the discriminators twice for every step of the generator with Adam optimizers. A batch size of 6 is used on a Tesla V100 GPU, with each input of 32000 samples.
The audio samples for the experiments are available at our project website:\\
{\tt{\footnotesize{https://pixl.cs.princeton.edu/pubs/Su\_2020\_HiFi/}}}

\subsection{Joint speech enhancement}
In joint speech enhancement experiment for conversion from real-world recording to clean studio-quality recording, our training set is constructed via data simulation. We use the studio-quality speech recordings from the Device and Produced Speech (DAPS) Dataset's clean set~\cite{mysore2015can}, convolve with the 270 impulse responses of the MIT Impulse Response Survey Dataset~\cite{traer2016statistics}, and then add noise drawn from the REVERB Challenge
database~\cite{kinoshita2013reverb} and the ACE Challenge database~\cite{eaton2016estimation}. In total, the data includes two genders $\times$ ten speakers per gender $\times$ ten minutes of script~$\times$ 270  environments before augmentation. We hold out one speaker for each gender, two minutes of script and 70 environments for evaluation.
The DAPS Dataset also provides recordings of the same set of studio-quality speech re-recorded under twelve different room environments, thus capturing interactions of acoustic factors in real world. This enables us to also test out how well our trained model generalizes to real unseen environments.

We conduct an ablation study by adding one building block at a time. The building blocks of our method include: feed-forward WaveNet (\textbf{Base}), use of the postnet (\textbf{Postnet}), use of the spectrogram discriminator (\textbf{SpecGAN}), and use of the waveform discriminators (\textbf{WaveGAN}). We also compare to six state-of-the-art methods in denoising and dereverberation: joint WPE~\cite{nakatani2010speech} and Wiener Filtering~\cite{scalart1996speech} (\textbf{W+W}), Deep Feature Loss for denoising~\cite{germain2019speech} (\textbf{Deep FL}), Attention Wave-U-Net~\cite{giri2019attention} (\textbf{Wave-U-Net}), spectral masking with Bidirectional LSTM~\cite{mack2018single} (\textbf{BLSTM}), and \textbf{MetricGAN}~\cite{fu2019metricgan}. 

\subsubsection{Objective evaluations}
Table~\ref{tab:Synthetic} reports objective evaluations on both the synthetic test set and the DAPS Dataset: Perceptual Evaluation of Speech Quality (PESQ), Short-Time Objective Intelligibility (STOI), Speech-to-reverberation Modulation Energy Ratio (SRMR), and Frequency-weighted Segmental SNR (FW-SSNR), which are commonly used metrics in denoising and dereverberation tasks, e.g. the 2014 REVERB Challenge~\cite{kinoshita2013reverb}.

\newcommand{\ignorethis } [1] {}

\ignorethis{
%
\begin{table}[th]
  \caption{Objective measures on the synthetic test set.}
 \vspace{-0.25\baselineskip}
  \label{tab:Synthetic}
  \centering
  \resizebox{0.99\linewidth}{!}{
  \begin{tabular}{l c c c c}
    \toprule
    \multicolumn{1}{l}{\textbf{Method}} & \multicolumn{1}{c}{\textbf{PESQ}} & \multicolumn{1}{c}{\textbf{STOI}} &
    \multicolumn{1}{c}{\textbf{SRMR}} & \multicolumn{1}{c}{\textbf{FW-SSNR}} \\
    \toprule
    \textbf{Clean} & 4.64 & 1.00 & 7.82 & 35.00\\
    \textbf{Noisy} & 1.92 & 0.91 & 5.64 & 4.63 \\
    \midrule
    \textbf{WPE+Wiener~\cite{nakatani2010speech, scalart1996speech}} & 2.20 & 0.90 & 6.81 & 4.53 \\
    \textbf{Deep FL~\cite{germain2019speech}} & 2.01 & 0.88 & 6.39 & 7.02 \\
    \textbf{Wave-U-Net~\cite{giri2019attention}} &  2.01 & 0.94 & 7.59 & 8.12 \\
    \textbf{BLSTM~\cite{mack2018single}} & 2.12 & 0.93 & 7.01 & 9.57
    \\
    \textbf{MetricGAN~\cite{fu2019metricgan}} & 2.59 & 0.92 & 7.28 & 6.23 \\ 
    \midrule
    \textbf{Base} & 2.50 & 0.95 & 7.16 & 11.52\\
    \textbf{+Postnet} & 2.60 & {\bf 0.95} & 7.48 & \textbf{11.68} \\
    \textbf{+Postnet +SpecGAN} & 2.69 & 0.94 & \textbf{7.83} & 10.24 \\
    \textbf{+Postnet +WaveGAN} & 2.56 & 0.95 & 7.62 & 10.95 \\
    \textbf{HiFi-GAN (All)} & \textbf{2.78} & 0.94 & 7.47 & 10.52 \\
    \bottomrule
  \end{tabular}
 }
\end{table}

\begin{table}[th]
\vspace{-0.6\baselineskip}
  \caption{Objective measures on the DAPS Dataset.}
  \vspace{-0.3\baselineskip}
  \label{tab:DAPS}
  \centering
  \resizebox{0.99\linewidth}{!}{
  \begin{tabular}{l c c c c}
    \toprule
    \multicolumn{1}{l}{\textbf{Method}} & \multicolumn{1}{c}{\textbf{PESQ}} & \multicolumn{1}{c}{\textbf{STOI}} &
    \multicolumn{1}{c}{\textbf{SRMR}} & \multicolumn{1}{c}{\textbf{FW-SSNR}} \\
    \toprule
    \textbf{Clean} & 4.64 & 1.00  & 7.82 & 35.00 \\
    \textbf{Noisy} & 1.41 & 0.87  & 4.79 & 3.04  \\
    \midrule
    \textbf{WPE+Wiener~\cite{nakatani2010speech, scalart1996speech}} & 1.84 & 0.87 & \textbf{7.84} & 3.61 \\
    \textbf{Deep FL~\cite{germain2019speech}} & 1.63 & 0.85 & 6.96 & 5.92 \\  
    \textbf{Wave-U-Net~\cite{giri2019attention}}  & 1.47 & 0.86 & 6.58 & 4.70 \\
    \textbf{BLSTM~\cite{mack2018single}} & 1.63 & 0.88 & 6.65 & 6.61 \\
    \textbf{MetricGAN~\cite{fu2019metricgan}}  & 1.89 & 0.88 & 7.38 & 4.73 \\ 
    \midrule
    \textbf{Base} & 1.76 & 0.89 & 6.80 & 6.38\\
    \textbf{+Postnet} & 1.93 & \textbf{0.89} & 7.22 & 7.44 \\
    \textbf{+Postnet +SpecGAN} & 1.97 & 0.87 & 7.46 & 7.44 \\
    \textbf{+Postnet +WaveGAN} & 1.86 & 0.88 & 7.48 & 6.52 \\
    \textbf{HiFi-GAN (All)} & \textbf{2.00} & 0.89 & 7.67 & \textbf{7.62} \\
    \bottomrule
  \end{tabular}
 }
 \vspace{-1.25\baselineskip}
\end{table}

} 

\begin{table}[th]
  \caption{Objective measures on syntheic and DAPS datasets.}
 \vspace{-0.25\baselineskip}
  \label{tab:Synthetic}
  \centering
  \resizebox{0.99\linewidth}{!}{
  \begin{tabular}{l c c c c}
    \multicolumn{1}{l}{\textbf{Method}} & \multicolumn{1}{c}{\textbf{PESQ}} & \multicolumn{1}{c}{\textbf{STOI}} &
    \multicolumn{1}{c}{\textbf{SRMR}} & \multicolumn{1}{c}{\textbf{FW-SSNR}} \\
    \bottomrule
    \vspace*{\baselineskip}\\
    \multicolumn{5}{c}{\textbf{Synthetic Data}} \\
    \toprule
    \textbf{Clean} & 4.64 & 1.00 & 7.82 & 35.00\\
    \textbf{Noisy} & 1.92 & 0.91 & 5.64 & 4.63 \\
    \midrule
    \textbf{WPE+Wiener~\cite{nakatani2010speech, scalart1996speech}} & 2.20 & 0.90 & 6.81 & 4.53 \\
    \textbf{Deep FL~\cite{germain2019speech}} & 2.01 & 0.88 & 6.39 & 7.02 \\
    \textbf{Wave-U-Net~\cite{giri2019attention}} &  2.01 & 0.94 & 7.59 & 8.12 \\
    \textbf{BLSTM~\cite{mack2018single}} & 2.12 & 0.93 & 7.01 & 9.57
    \\
    \textbf{MetricGAN~\cite{fu2019metricgan}} & 2.59 & 0.92 & 7.28 & 6.23 \\ 
    \midrule
    \textbf{Base} & 2.50 & 0.95 & 7.16 & 11.52\\
    \textbf{+Postnet} & 2.60 & {\bf 0.95} & 7.48 & \textbf{11.68} \\
    \textbf{+Postnet +SpecGAN} & 2.69 & 0.94 & \textbf{7.83} & 10.24 \\
    \textbf{+Postnet +WaveGAN} & 2.56 & 0.95 & 7.62 & 10.95 \\
    \textbf{HiFi-GAN (All)} & \textbf{2.78} & 0.94 & 7.47 & 10.52 \\
    \bottomrule
    \vspace*{\baselineskip}\\
    \multicolumn{5}{c}{\textbf{DAPS Dataset}} \\
    \toprule
    \textbf{Clean} & 4.64 & 1.00  & 7.82 & 35.00 \\
    \textbf{Noisy} & 1.41 & 0.87  & 4.79 & 3.04  \\
    \midrule
    \textbf{WPE+Wiener~\cite{nakatani2010speech, scalart1996speech}} & 1.84 & 0.87 & \textbf{7.84} & 3.61 \\
    \textbf{Deep FL~\cite{germain2019speech}} & 1.63 & 0.85 & 6.96 & 5.92 \\  
    \textbf{Wave-U-Net~\cite{giri2019attention}}  & 1.47 & 0.86 & 6.58 & 4.70 \\
    \textbf{BLSTM~\cite{mack2018single}} & 1.63 & 0.88 & 6.65 & 6.61 \\
    \textbf{MetricGAN~\cite{fu2019metricgan}}  & 1.89 & 0.88 & 7.38 & 4.73 \\ 
    \midrule
    \textbf{Base} & 1.76 & 0.89 & 6.80 & 6.38\\
    \textbf{+Postnet} & 1.93 & \textbf{0.89} & 7.22 & 7.44 \\
    \textbf{+Postnet +SpecGAN} & 1.97 & 0.87 & 7.46 & 7.44 \\
    \textbf{+Postnet +WaveGAN} & 1.86 & 0.88 & 7.48 & 6.52 \\
    \textbf{HiFi-GAN (All)} & \textbf{2.00} & 0.89 & 7.67 & \textbf{7.62} \\
    \bottomrule
  \end{tabular}
 }
 \vspace{-1.25\baselineskip}
\end{table}

We found that the Deep Feature Loss network trained on our dataset produces worse results than the pre-trained model, likely because its loss network is attending additive environmental noise but not reverberation differences. Therefore, we run the pre-trained Deep Feature Loss model for comparison. 

Our HiFi-GAN method achieves the best PESQ among all in both cases. However, the use of the waveform discriminator alone can degrade the scores. We hypothesize that having the spectrogram discriminator alongside the waveform discriminators stabilizes the adversarial training, thus leading to a better convergence point for HiFi-GAN. We also observe that the waveform-based methods generally do not work as well as spectral ones when reverberation is present. This shows importance to include time-frequency representations into learning. 

There is no consistent ranking across the four objective metrics, suggesting that no single measure adequately captures the subjective sense of the overall quality, also observed by previous works~\cite{kinoshita2016summary}. Therefore, we run subjective evaluations for human judgement of overall perceptual qualities.
\begin{figure}[t]
  \centering
  \includegraphics[width=0.95\linewidth]{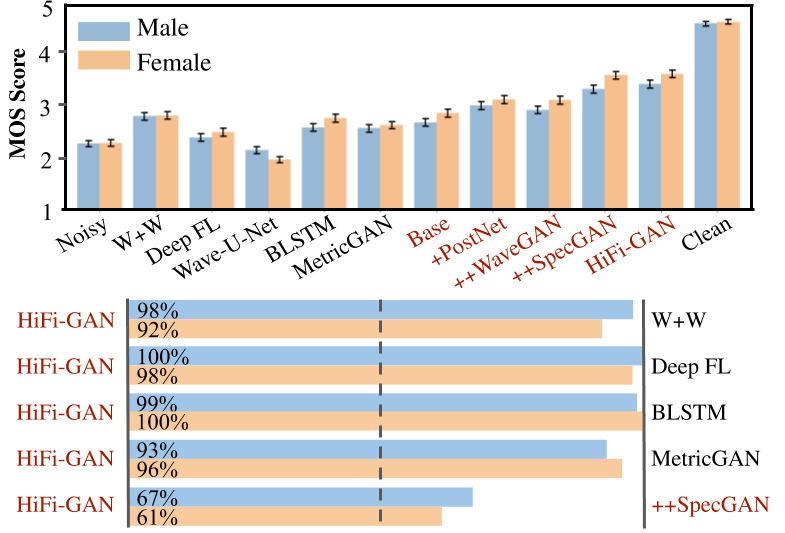}
  \vspace{-0.1in}
  \caption{MOS scores (top) and percentage of preference on HiFi-GAN in pairwise study (bottom) on the DAPS Dataset.}
  \label{fig:MOS}
  \vspace{-1.5\baselineskip}
\end{figure}

\subsubsection{Subjective evaluations}
We use Amazon Mechanical Turk (AMT) for listening study. In a Mean Opinion Score (MOS) test, a subject is asked to rate the sound quality of an audio snippet on a scale of 1 to 5, with 1=\emph{Bad}, 5=\emph{Excellent}. We provide a studio-quality audio as reference for high-quality, and the input noisy audio as low-anchor.
In total, we collected about 1000 valid HITs per voice, totalling 14K ratings per voice. As is shown in Figure~\ref{fig:MOS}, HiFi-GAN scores highest, and all of our variants are among the best. While MetricGAN seems competitive in objective measurement, our methods significantly outperform it with a p-value $<10^{-4}$.
%
Since MOS ratings are relatively close for a few methods, we also conduct an preference test on the selected method pairs to reveal the consistency of quality of our top-performing HiFi-GAN versus competitors. In this study, a subject is presented with two utterances produced by two methods and asked to choose which one sounds better given a clean recording as reference. After all the answers are collected, we assign the preferred method to each utterance based on majority voting to reduce high variation in ratings. Each method pair received 900 ratings per voice. Our method is strongly preferred over other baselines (more than 90\% of the times), and HiFi-GAN shows evident advantage over SpecGAN which scores almost equally in MOS test.
%

\subsection{Denoising task}
To further demonstrate the effectiveness of HiFi-GAN, we conducted a benchmark speech enhancement (denoising) experiment, using a standard dataset~\cite{valentini2016investigating} consisting of pairs of clean and noisy recordings (28 speakers for training and 2 speakers for test). The objective measures of our method in comparison to other state-of-the-art methods are reported in Table~\ref{tab:VCTK}.

HiFi-GAN outperforms all the other methods, and Postnet is second only to MetricGAN. This shows our approach can generalize to different speech enhancement tasks and datasets. 
Note that the clean recordings from the VCTK Dataset still contain a small amount of background noise, and thus our methods score lower on the background distortion measure (CBAK).

\begin{table}[th]
 \vspace{-0.5\baselineskip}
  \caption{Objective measures on the VCTK Noisy Dataset}
  \vspace{-0.5\baselineskip}
  \label{tab:VCTK}
  \centering
   \resizebox{0.85\linewidth}{!}{
  \begin{tabular}{c c c c c}
    \toprule
    \multicolumn{1}{c}{\textbf{Method}} & \multicolumn{1}{c}{\textbf{PESQ}} & \multicolumn{1}{c}{\textbf{CSIG}} &
    \multicolumn{1}{c}{\textbf{CBAK}} & \multicolumn{1}{c}{\textbf{COVL}} \\
    \midrule
    \textbf{Noisy} & 1.97 & 3.35 & 2.44 & 2.63  \\
    \textbf{Wiener~\cite{scalart1996speech}}  & 2.22 & 3.23 & 2.68 & 2.67 \\ 
    \textbf{SEGAN~\cite{pascual2017segan}}   & 2.16 & 3.48 & 2.94 & 2.80  \\
    \textbf{WaveNet~\cite{rethage2018WaveNet}} & - & 3.62 & 3.23 & 2.98 \\
    \textbf{Deep FL~\cite{germain2019speech}} & - & 3.86 & 3.33 & 3.22 \\
    \textbf{Wave-U-Net~\cite{giri2019attention}}  & 2.62  & 3.91 & \textbf{3.35} & 3.27 \\
    \textbf{MetricGAN~\cite{fu2019metricgan}}  & 2.86 & 3.99 & 3.18 & 3.42   \\
    \midrule
    \textbf{Postnet}  & 2.84 & \textbf{4.18} & 2.55 & \textbf{3.51} \\
    \textbf{HiFi-GAN}  & \textbf{2.94}  & 4.07 & 3.07 & 3.49  \\
    \bottomrule
  \end{tabular}
  }
  \vspace{-1.5\baselineskip}
\end{table}

%




\section{Discussion and Future Work}
 \label{sec:discussion}
In this paper, we present HiFi-GAN, an end-to-end deep learning method for enhancing speech recordings to studio-quality. It uses a feed-forward WaveNet together with multi-scale adversarial training in both time domain and time-frequency domain. The dynamic matching losses of deep features of the discriminators help to achieve better perceptual qualities. Evaluations show the proposed method outperforms all the other state-of-the-art baselines in denoising and dereverberation tasks.

Note that our experiments are conducted on sample rate of 16kHz, 
but real high-fidelity should target at 44kHz. This can be achieved by retraining the networks on the desired sample rate, which is however computationally expensive. Alternatively, one potential future work is to attach a band extension network~\cite{feng2019learning} for postnet to up-sample from 16kHz to 44kHz.

\bibliographystyle{IEEEtran}
\bibliography{paper}

\end{document}